\documentclass[printer]{aa}

\newcommand{\out}[1]{}  

\newcommand{\map}{\hbox{{\sc mappings i}c}}
\newcommand{\up}{\hbox{\it U}}

\newcommand{\mic}{\hbox{$\mu{\rm m}$}}
\newcommand{\cms}{\hbox{${\rm cm^{-2}}$}}

\newcommand{\kms}{\hbox{${\rm km\,s^{-1}}$}}

\newcommand{\llu}{\hbox{${\rm erg\, cm^{-2}\, s^{-1}}$}}
\newcommand{\lu}{\hbox{${\rm erg\, cm^{-2}\, s^{-1}}\,{\rm \AA}^{-1}$}}

\newcommand{\ad}{\hbox{$\delta$}}
\newcommand{\dc}{\hbox{$\delta_{\rm C}$}}

\newcommand{\anu}{\hbox{$\alpha_{\nu}$}}

\newcommand{\taum}{\hbox{$\tau_{640}$}}
\newcommand{\al}{\hbox{$\alpha$}}
\newcommand{\aox}{\hbox{$\alpha_{OX}$}}

\newcommand{\zq}{\hbox{$z_q$}}

\newcommand{\I}{\hbox{{\sc i}}}
\newcommand{\II}{\hbox{\sc ii}}
\newcommand{\III}{\hbox{\sc iii}}
\newcommand{\IV}{\hbox{\sc iv}}

\newcommand{\HI}{\hbox{H\,{\sc i}}}

\newcommand{\HeI}{\hbox{He\,{\sc i}}}

\newcommand{\OVI}{\hbox{O\,{\sc vi}}}

\newcommand{\NII}{\hbox{N\,{\sc ii}}}

\newcommand{\NV}{\hbox{N\,{\sc v}}}
\newcommand{\CIV}{\hbox{C\,{\sc iv}}}
\newcommand{\CV}{\hbox{C\,{\sc v}}}

\newcommand{\w}{\hbox{$\lambda$}}
\newcommand{\OVIw}{\hbox{O\,{\sc vi}\,$\lambda 1035$}}
\newcommand{\NVw}{\hbox{N\,{\sc v}\,$\lambda 1240$}}

\newcommand{\CIVw}{\hbox{C\,{\sc iv}\,$\lambda 1549$}}

\newcommand{\HE}{\hbox{HE\,2347$-$4342}}

\newcommand{\ton}{\hbox{Ton\,34}}

\newcommand{\Lya}{\hbox{Ly$\alpha$}}

\newcommand{\Nh}{\hbox{$N_{\rm H}$}}
\newcommand{\NHI}{\hbox{$N_{\rm HI}$}}

\newcommand{\NN}{\hbox{$N_{20}^{\rm H}$}}
\newcommand{\NC}{\hbox{$N_{\rm C}$}}
\newcommand{\NCdust}{\hbox{$N_{\rm C.dust}$}}
\newcommand{\NCgas}{\hbox{$N_{\rm C.gas}$}}
\newcommand{\NCIV}{\hbox{$N_{\rm CIV}$}}

\newcommand{\Cut}{\hbox{$C_{\lambda}$}}

\newcommand{\labr}{\hbox{$\lambda_{brk}$}}
\newcommand{\sed}{\hbox{{\sc sed}}}
\newcommand{\seds}{\hbox{{\sc sed}{\rm s}}}
\newcommand{\Fnu}{\hbox{F$_\nu$}}

\newcommand{\Fla}{\hbox{F$_\lambda$}}

\usepackage{graphics}

\begin{document}



\title{The unusual UV continuum of quasar Ton\,34 \\and the possibility of crystalline dust absorption}


\author{Luc Binette\inst{1,2} and Yair Krongold\inst{1}
          }

\authorrunning{Binette et~al.}


\institute{Instituto de Astronom\'\i a, UNAM,  Ap. 70-264, 04510
M\'exico, DF, M\'exico \and D\'{e}partement de Physique, de G\'{e}nie
Physique et d'Optique, Universit\'{e} Laval, Qu\'{e}bec, QC, G1K\,7P4}
\date{Received June 2007/ Accepted November 2007}

\titlerunning{The unusual UV \sed\ of \ton}

\authorrunning{Binette  \& Krongold}

\abstract{Luminous quasars are known to display a sharp steepening
of the continuum near 1100\,\AA. This spectral feature is not well
fitted by current accretion disk models, unless comptonization of
the disk emission is invoked. Absorption by carbon crystalline dust
has been proposed to account for this feature.} {\ton\ ($\zq=1.928$)
exhibits the steepest far-UV decline ($\Fnu \propto \nu^{-5.3}$)
among the 183 quasar HST-FOS spectra analyzed by Telfer et\,al. It
is an ideal object to test the crystalline dust hypothesis as well
as alternative interpretations of the UV break.}{We reconstruct the
UV spectral energy distribution of \ton\ by combining HST, IUE and
Palomar spectra.} {The far-UV continuum shows a  very deep continuum
trough, which is bounded by a steep far-UV rise. We fit the trough
assuming nanodiamond dust grains.}{Extinction by carbon crystalline
dust reproduces the deep absorption trough of \ton\ reasonably well,
but not the observed steep rise in the extreme UV. We also study the
possibility of an intrinsic continuum rollover. The dust might be
part of a high velocity outflow ($\simeq 13000$\,\kms), which is
observed in absorption in the lines of \CIVw, \OVIw, \NVw\ and
\Lya.}


\maketitle

\section{Introduction} \label{sec:intro}

The spectral energy distribution (\sed) of  quasar and Seyfert\,I
galaxies is composed of emission lines superimposed on various
continuum emission components. The near-infrared to visible domain
is reasonably well reproduced by a powerlaw. The ultraviolet
spectral region is characterized by a broad continuum excess, which
is referred to as the big blue bump (BBB). According to general
belief, it corresponds to the thermal signature from a hot accretion
disk orbiting a supermassive black hole. The extension in the
extreme UV of the BBB is expected to provide the ionizing flux that
powers most of the emission lines. A serious problem with this
picture, however, is that the BBB appears to be too
soft\footnote{There is an apparent inconsistency between the \seds\
typically observed and the much harder ones preferred in
photoionization BELR models (e.g. Baldwin et\,al. 1995; Casebeer,
Leighly \& Baron 2006).} to account for the high excitation lines
from the broad emission line region (BELR) (Korista et\,al. 1997).
In effect, a marked continuum decline (i.e. steepening) takes place
shortward of $\simeq 1100$\,\AA\ (rest-frame), which we hereafter
refer to as the far-UV break\footnote{We will also refer to the
wavelength domain longward and shortward of the \sed\ steepening (or
break) as the near- and far-UV regions, respectively.}. A
possibility might be that this break is more akin to a localized
continuum trough, followed by a marked recovery in the extreme UV,
which is the energy region responsible for producing the
high-excitation emission lines. Various mechanisms that could
generate such a trough are summarized by Binette et\,al. (2007).


Our aim is to probe the nature of the far-UV break and to test
alternative interpretations of it by focussing on the more extreme
cases. For instance, many quasar spectra reveal an \sed\
significantly steeper than the `average' \sed\ derived by Telfer
et\,al.( 2002; hereafter TZ02), which behaves as $\nu^{-1.76}$ (in
$F_{\nu}$) shortward of the break. Among the 77 far-UV indices
measured by TZ02, there were 3 objects with an ionizing continuum
steeper than $\nu^{-3}$. In this respect, \ton\ (alternatively named
PG\,1017+280 or J1019+2745) at redshift $\zq=1.928$ is the most
extreme case, with a powerlaw index as steep as $\nu^{-5.3}$ (TZ02).
For this reason, we consider it an ideal object to test competing
models of the physical origin of the 1100\,\AA\ break. In this
paper, we used different bibliographical sources to build an \sed\
as complete as possible of \ton.
As the data are of limited quality, we intend to obtain higher
quality data that would cover the X-ray domain down to the optical
UV.

Our initial objective was to verify whether the hypothesis of
absorption by crystalline carbon dust grains of Binette et\,al.
(2005; hereafter B05) would survive the test of modeling the extreme
\sed\ of \ton. Using two flavors of nanodiamonds, B05 could
successfully reproduce the position and detailed shape of the far-UV
break in 50 quasars, out of a total sample of 61 objects from
HST-FOS archives\footnote{The analysis of B05 included only  {\it
multigrating} HST spectra that extended down to at least 900\,\AA\
(rest-frame), i.e. 61 quasars in total.  \ton\ was not included,
since only a single grating HST spectrum exists.}. More recently,
Haro-Corzo et\,al. (2007, hereafter H07) reduced the dust model to a
single flavor (that excludes the meteoritic type), since a
particular emission feature expected near 3.5\,$\mu$m with the
meteoritic case is absent from the mid-IR spectrum of 3C298 (de
Diego et\,al. 2007).

In this paper, we discuss the merits of the dust absorption model as
well as its limitations in reproducing the deep trough that
characterize the \ton\ \sed. In Appendix\,\ref{app:a}, we compare
different extinction curves and illustrate how the far-UV extinction
is affected. In a companion Paper (Binette \& Krongold 2007), we
report on some of the peculiarities of the emission line spectrum of
this unusual quasar.

\begin{figure}
\resizebox{\hsize}{!}{\includegraphics{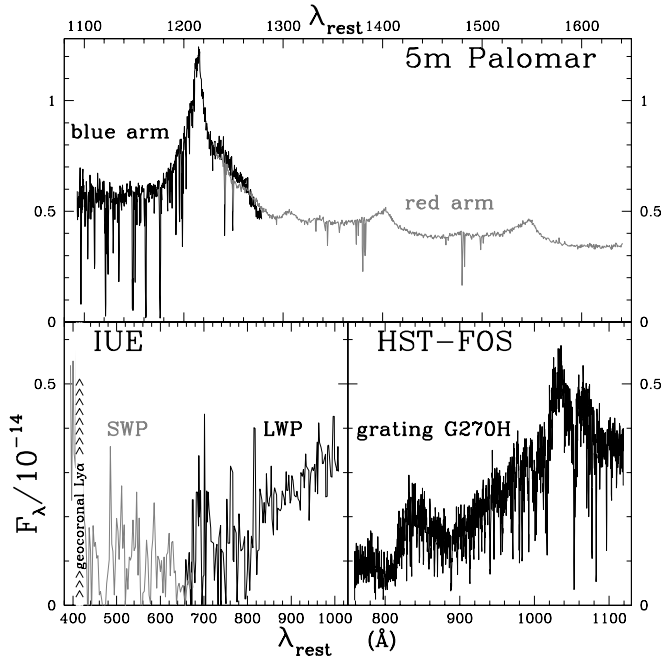}}
\caption{Spectral segments of \ton\ extracted from different
archival or bibliographical sources. Top panel: blue and red optical
spectra (in arbitrary units) from Sargent, Boksenberg \& Steidel
(1988); lower right panel: HST-FOS spectrum using grating G270H,
lower left panel: IUE spectra (LWP and SWP) (see
Sect.\,\ref{sec:spc}). \label{fig:spc}}
\end{figure}

\section{The observed UV energy distribution of \ton}\label{sec:obs}

\subsection{Description of the archival data}\label{sec:spc}

The current work is based on the following  archival or
bibliographical sources. The 760--1120\,\AA\ spectral segment is
provided by the dataset Y2IE0A0AT from the HST-FOS archives. It
corresponds to an integration time of 1100\,s (23rd of February
1995) using grating G270H (see lower right panel in
Fig.\,\ref{fig:spc}). To cover the extreme UV, we then borrowed from
the IUE archives. The long wavelength segment (LWP in
Fig.\,\ref{fig:spc}) is from Tripp, Bechtold \& Green (1994) and
corresponds to the dataset LW0P5708 (25800\,s on 18th of April
1986).  Fluxes longward of 3000\,\AA\ (observer-frame) were severely
affected from reflected sunlight or moonlight (Lanzetta, Turnshek \&
Sandoval 1993) and have been discarded. The shorter wavelength IUE
segment (SWP in Fig.\,\ref{fig:spc}) was extracted directly from the
archives and corresponds to the dataset SWP28188 (23400\,s on 4th of
October 1985). Because of the limited S/N and to avoid line clutter,
the SWP segment shown in Fig.\,\ref{fig:spc}  has been re-binned by
merging pixels in groups of 5. The region corresponding to the
strong geocoronal \Lya\ line has been masked.

In order to constrain the \sed\ behavior longward of the HST
segment, we adopted the published optical spectra of Sargent,
Boksenberg \& Steidel (1988), which were taken at the Palomar
5.08\,m Hale Telescope (in November 1981 and February 1982).  Both
optical spectra (blue and red arm segments in top panel of
Fig.\,\ref{fig:spc}) lacked absolute flux calibration. The authors
observed standard stars, which allowed them to provide a relative
calibration.

\subsection{Derivation of the UV \sed}  \label{sec:sed}

The ultraviolet \sed\ from \ton\ was derived in the following
manner: we statistically corrected the UV spectral segments for the
cumulated absorption caused by unresolved \Lya\ forest lines,  which
are responsible of  the so-called far-UV ``Lyman valley'' (M{\o}ller \&
Jakobsen 1990). For that purpose, we adopted the
\emph{mean}\footnote{This correction is statistical in nature, as it
relies on the average behavior with redshift of the spatial density
of intervening absorbers. It cannot be used to correct small
portions of the continuum, which may be a coincident with a ``clear
patch'' or an over-density in the \Lya\ forest. These
inhomogeneities may generate spurious narrow features that  should
not be  attributed to emission lines.} transmission function for
$\zq=2$ published by Zheng et\,al. (1997). We also applied a
Galactic reddening correction assuming the Cardelli, Clayton \&
Mathis (1989) extinction curve corresponding to ${\rm R}_{\rm
V}=3.1$ and ${\rm E}_{B-V} = 0.13$. The latter value corresponds to
the mean extinction inferred  from the 100$\,\mu$ maps of Schlegel
et\,al. (1998) near \ton.  The blue and red arm segments have been
scaled to overlap smoothly with the HST-FOS segment.

Both the LWP and SWP segments were multiplied by a factor 0.75. This
scaling was necessary so that the LWP segment superimposes the
HST-FOS spectrum as closely as possible.  Continuum variability is a
possible explanation for this continuum difference, since the IUE
and HST observations were made in different years.

To derive the UV \sed\ shown in Fig.\,\ref{fig:ton}, all the
spectral segments were shifted to rest-frame wavelengths, and \Fla\
was multiplied by $1+\zq$. The IUE spectra have been re-binned by
grouping $n$ pixels together (SWP with $n=5$ and LWP with $n=3$) in
order to improve the limited S/N and to avoid overcrowding due to
line cluttering. The different spectral segments have been color
coded as follows, SWP: red, LWP orange, HST-FOS: blue, and Palomar:
dark green.

\subsection{Originality and limitations of the data}\label{sec:lim}

The far-UV HST segment is characterized by a very sharp drop. The
spectral index is as steep as  $-5.3$ according to TZ02. We
emphasize that both the LWP and the HST data confirm the existence
of a sharp flux decline. The \sed\ as a whole suggests the existence
of a very deep trough, which reaches its lowest point at $\sim
650$\,\AA, followed by a flux rise shortward of the \Lya\ geocoronal
line, as indicated by the SWP segment.

How can we explain the existence  of such a deep trough in the
far-UV? If it was due to the Lyman Valley, it would imply an
increase of a factor of more than 5 in the density of \Lya\ forest
absorbers. This excess would have to extend along the line of sight
over a redshift span of order unity, which appears to be very
unlikely. The existence of a few thick absorbers is another
possibility. However, such absorbers would result in sharp saturated
absorption lines, unlike the progressive drop in flux observed in
both the LWT and the HST spectra.

In our opinion, the deep trough seen in \ton\ is a manifestation of
the far-UV break commonly observed in quasars, albeit in a more
extreme form. If the steep flux rise seen in \ton\ towards 400\,\AA\
was confirmed, it would lend support to the absorption hypothesis
presented in Sect.\,\ref{sec:ext}. Alternative explanations could
also be explored (c.f. Binette et\,al. 2007).

As can be gathered from Fig.\,\ref{fig:ton}, the strongest emission
features in the far-UV coincide with the position of lines observed
or expected in quasar spectra. The strengths of some of these lines
is unusual and is the subject of a separate paper (Binette \&
Krongold 2007).

\section{Dust extinction models} \label{sec:ext}

\subsection{Carbon crystallite extinction} \label{sec:cur}

TZ02 suggested that dust might play a role in the three quasars that
presented the steepest far-UV indices. B05 found that only by
considering crystallite carbon grains could they reproduce those
\seds, which had the most pronounced far-UV breaks (the so-called
class\,B spectra in B05), because the extinction in that case is
characterized by a relatively sharp threshold in the UV. Since \ton\
shows the most extreme class\,B spectrum, we adopt the corresponding
extinction curve D1 from B05, which was computed assuming a powerlaw
size distribution ($a^{-3.5}$)  of spherical grains consisting of
sizes ranging between 3 and 25\,\AA. The grain composition
corresponds to cubic nanodiamonds\footnote{Instead of using two
types of nanodiamonds (cubic and meteoritic) as in B05, H07 could
fit the more numerous class\,A \seds\ using only cubic nanodiamonds
(without the meteoritic type), but with a much wider grain size
distribution that extended from 3 to 200\,\AA.}  (i.e. {\it without}
surface adsorbates).

A review of the advantages of dust grains consisting of nanodiamonds
is presented in Appendix\,\ref{app:a}. We do not consider the
nanodiamond hypothesis as the final answer  (see Binette et\,al.
2007), but as one possibility that warrant further study. For the
sole purpose of procuring a convenient normalization of the
extinction cross section of the dust model labeled D1 (shown in
Fig.\,\ref{fig:comp}), all carbon was assumed to be depleted onto
dust, with a solar carbon abundance (this assumption is relaxed in
Sect.\,\ref{sec:out}).

The transmitted flux across the dusty material is given by
$F_{\lambda}^{obs}= T_{\lambda} F_{\lambda}^q$, where $T_{\lambda}$
is the transmission function, which we approximate with an
exponential $e^{-\tau_{\lambda}^{ext}}$.  Since the albedo for the
D1 extinction curve is negligible, scattering need not be considered
for the transfer. The opacity is $\tau_{\lambda}^{ext} = \Nh
\sigma_{\lambda}^{H}$, where $\sigma_{\lambda}^{H}$ is that given by
curve D1 and $\Nh$, the H column density,  is a free parameter
determined by the  fitting  of the UV trough. Hereafter, we will use
$\NN$, which is the H column in units of $ 10^{20} \,\cms$.  As a
rule of thumb (and coincidentally), the peak opacity at 640\,\AA\ is
simply given by \NN\ ($\simeq \taum$). We will assume that the
intrinsic \sed\ of \ton\ follows a powerlaw in the near-UV, for
which the index \anu\ and normalization constant $\cal{B}$ are
defined as follows:
\begin{eqnarray} \label{eq:pow}
\begin{array}{rl}
  \,&F_{\lambda}^q = \cal{B} \, (\frac{\lambda}{\lambda_{\rm r}})^{-({\rm 2}+\anu)} ,
\end{array}
\end{eqnarray}
where $\lambda_r = 1610\,$\AA\ is used as a reference wavelength. As
is customary, the \sed\ indices are defined in the plane \Fnu\
($\propto \nu^{+\alpha_{\nu}}$).

\begin{figure}
\resizebox{\hsize}{!}{\includegraphics{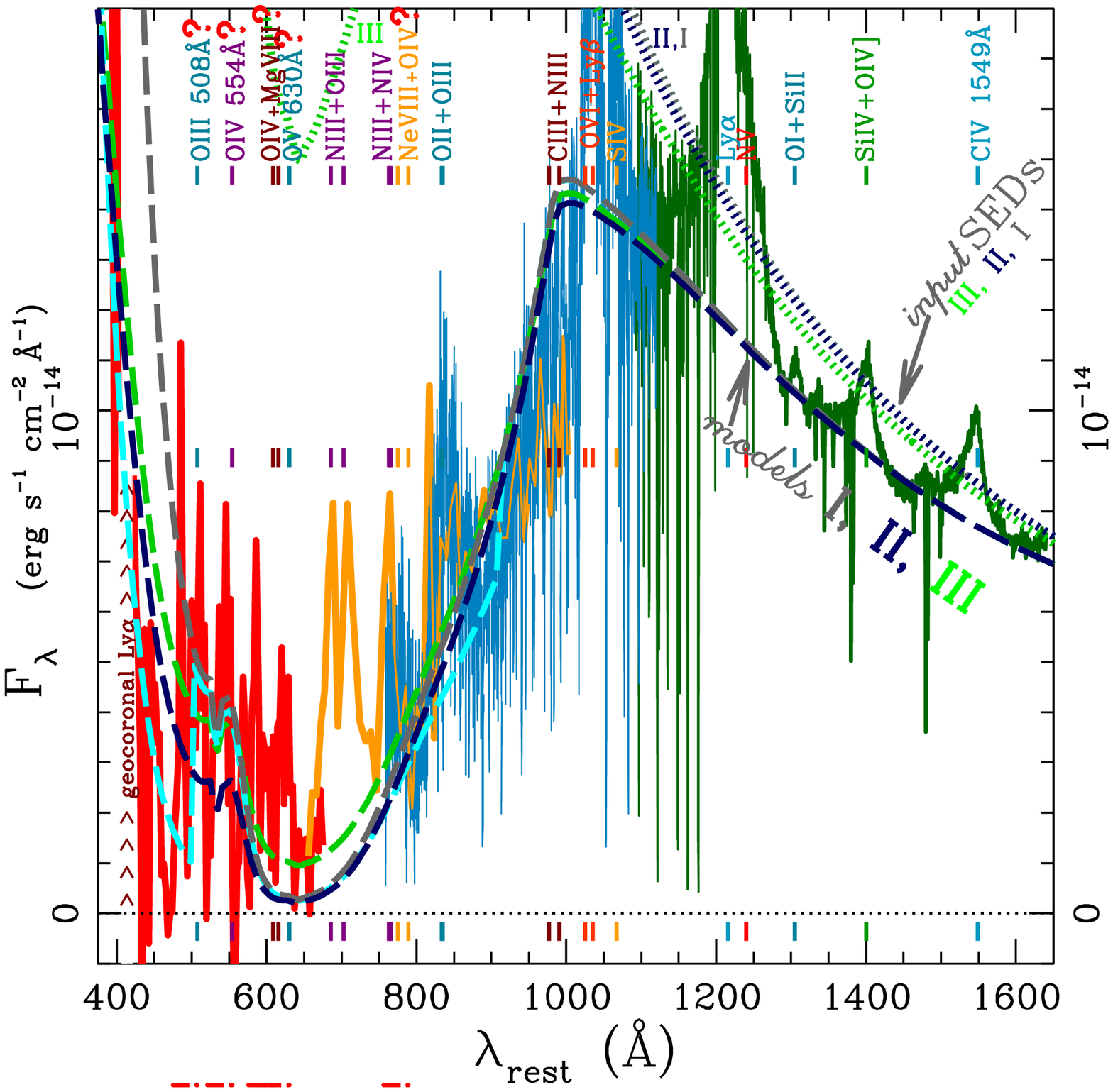}}
\caption{{The UV spectral energy distribution in \Fla\ (\lu) of
\ton\ ($\zq=1.928$) as a function of rest-frame wavelength.  The
near-UV spectrum (dark green continuous line) is from Sargent,
Boksenberg \& Steidel (1988) and has been scaled to overlap the
HST-FOS spectrum (blue line). The far-UV segments correspond to the
LWP (orange line) and SWP (red line) spectra from IUE, and have been
scaled by a factor 0.75 (see Sect.\,\ref{sec:sed}). The whole \sed\
has been corrected for Galactic reddening (${\rm E}_{B-V} = 0.13$)
and for the cumulative absorption by unresolved intergalactic \Lya\
forest lines. The geocoronal \Lya\ emission line has been masked.
Color-coded fiducial marks indicate the position of observed or
expected (labeled with symbol '?') emission lines. Dust absorption
Models\,\I, \II\ and \III\ overlay the observed spectrum using
color-coded {\it dashed} lines. The dotted lines of the
corresponding color represent the \emph{input} \sed\ before
absorption (these are reproduced in Log\,\Fla\ in
Fig.\,\ref{fig:sed}a).  Model\,\I\ (silver line): powerlaw \sed\
absorbed by a slab of column \NN=5.3. Model\,\II\ (navy blue line):
powerlaw \sed\ with a rollover at 670\,\AA\ absorbed by a slab of
column \NN=5.3. Model\,\III\ (color lime): same \sed,  first
absorbed by a slab with \NN=1.0 (dotted lime line is \sed\,\III),
then further  absorbed by a dust screen with thickness \NN=4.3 and a
leakage of 5\%. The cyan colored dashed line is the same as model
\I, except that it also considers the maximum column allowed for
absorption by  atomic \HI\ and \HeI, that is, $4\times 10^{16}$ and
$2\times 10^{17}$\,\cms, respectively (see Sect.\,\ref{sec:atom}).
All four dashed-line absorption models have been normalized  to the
same flux of ${\cal{B}} = 7.3 \times 10^{-15}\,$\lu\ at 1610\,\AA\
(see Eq.\,\ref{eq:pow}).} \label{fig:ton}}
\end{figure}

\subsection{Modeling the far-UV continuum, assuming dust
absorption}\label{sec:model}

\subsubsection{The simple powerlaw case} \label{sec:pow}

\begin{figure}
\resizebox{\hsize}{!}{\includegraphics{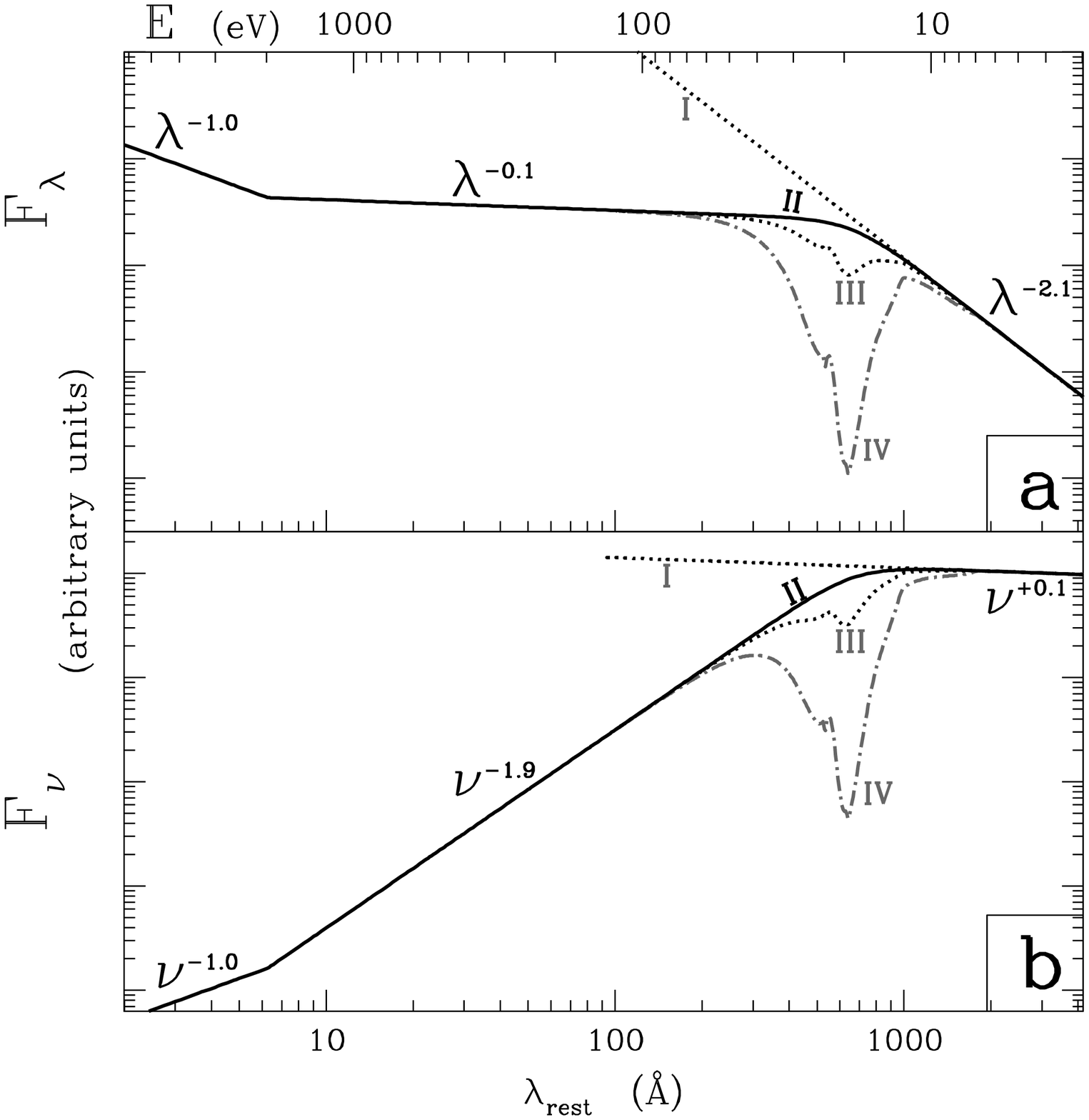}}
\caption{Log-log plot of the three input spectral energy
distributions \I\--\III\ in either \Fla\ (panel {\it a}) or \Fnu\
(panel {\it b}) as a function of wavelength (bottom axis) or photon
energy (top axis). The {\sl transmitted} \sed\ assuming $\NN=5.3$ is
shown by  the dash-dotted line,  labeled \IV\ above.
\label{fig:sed}}
\end{figure}

Longward of the break, the near-UV is well reproduced  by using an
index $\anu=-0.3$. Although small, the extinction within the near-UV
domain (for curve D1) cannot be neglected in the case of \ton.
Therefore, the assumed intrinsic \sed\ must be somewhat harder than
observed. We find that an index of $+0.1$ is favored by our dust
models. The dotted line labeled \I\ illustrates such a continuum in
both Figs.\,\ref{fig:ton} and \ref{fig:sed}. To reproduce the
trough, a dust screen of column \NN=5.3 is required. The resulting
fit is represented by the silver dashed line in Fig.\ref{fig:ton}.
The model presents a reasonable approximation of the through, but
the steep far-UV recovery occurs too early in this Model\,\I, that
is, longward of the observed rise.


\subsubsection{The case of a powerlaw with a far-UV roll-over}\label{sec:roll}

There are no X-ray observations reported for \ton. The source was
not detected by the ROSAT All Sky Survey (RASS). However, we find
that the extrapolation of the assumed powerlaw (see
Fig.\,\ref{fig:sed}) leads to a 0.1--2.4\,keV (observer-frame) flux
3--4 orders of magnitude larger than the RASS flux limit of $5\times
10^{-13}\,$\llu\ (Voges et\,al. 1999). Therefore, the near-UV
powerlaw must steepen considerably in the far-UV or in the soft
X-rays, but we have no direct indication as to where.  B05 proposed
the existence of a roll-over in the extreme UV, to connect the UV
domain with the X-rays. In their study of 11 individual quasar
\seds, H07 included the same roll-over. They parameterized it using
the function
\begin{eqnarray} \label{eq:cut}
\begin{array}{cc}
\Cut = \left(1+ \left[{\lambda}/{\lambda_{brk}}\right]^{f \delta}
\right)^{-f^{-1}}
\end{array}
\end{eqnarray}
where \ad\ ($<0$) is a measure of the steepening, $f$ a form factor
and \labr\ the wavelength of the cut-off. The values considered by
these authors are $f=2.8$ and \labr=670\,\AA\ (18.5\,eV). By
multiplying the near-UV powerlaw of Eq.\,\ref{eq:pow} by this
function \Cut, a shallow steepening now takes place at \labr, which
has the effect of incrementing the powerlaw index by \ad\ in the
far-UV. The value of \ad\ favored by B05 was $-1.6$. However, H07
showed that either the index change \ad\ is significantly larger, or
the roll-over must take place at significantly higher energies. For
\ton, we adopted a value of $\ad=-2.0$, justified as follows: the
\aox\ index, defined by the flux ratio between 2500\,\AA\ and
2\,keV, tends to fall in quasars in the interval $-1.5$ to $-1.6$
(Anderson \& Margon 1987; Green et\,al. 1995; Avni, Worrall \&
Morgan 1995; Yuan et\,al. 1998). Our chosen value of $\ad=-2.0$
results in an $\aox=-1.45$, which is consistent with these
determinations. It implies an X-ray flux of $\sim 6\times
10^{-13}\,$\llu\ between 0.1--2.4\,keV (observer-frame), consistent
with the RASS limit quoted above.

Such a \sed\ consisting of the powerlaw \I\ multiplied by \Cut\ is
labeled \II\ and shown in either panel of Fig.\,\ref{fig:sed}. For
the purpose of calculating photoionization models (Binette \&
Krongold 2007), we truncated \sed\,\II\ at 2\,keV and appended a
harder powerlaw to describe the hard X-ray domain. An index of
$\simeq -1.0$ better approximates the typical photon indices ($\sim
2$) measured in the 2--10\,keV domain (e.g. Williams et\, al. 1992;
Lawson et\,al. 1992).

The dust absorption model assuming \sed\,\II\ is represented by the
navy blue dashed line in Fig.\,\ref{fig:ton}. It has the same dust
screen column, \NN=5.3, as the previous model using \sed\,\I. The
fit to the steep far-UV rise has improved considerably with respect
to previous Model\,\I.

\subsubsection{The partially leaking case}\label{sec:leak}

We note that the observed trough shows  at its lowest point a
non-zero flux higher than that of Model\,\II. This flux is not
necessarily all continuum. It is possible that resonance lines such
as \NII\ at \w645\,\AA\ might be contributing near the bottom of the
trough. Assuming  that the minimum flux is all continuum, we
investigate the possibility of leakage of the screen as a result of
small inhomogeneities of the screen or of partial covering of the
source. As the opacity increases, the relative importance of leakage
increases relative to the transmitted flux, until it eventually
dominates when $\taum \gg 1$. To illustrate this, we have considered
a partially covered source and found that a leakage of order 5\%
suffices to reproduce the observed minimum flux.  This is
illustrated  by the lime dashed-line Model\,\III\ which represents
the transmitted+leaked flux from input \sed\,\III\ (see
Figs.\,\ref{fig:ton} and \ref{fig:sed}). This input \sed\,\III\
corresponds to \sed\,\II, but absorbed by a column of \NN=1.0
(without any leakage). The intention was to use an input \sed\ that
already reproduces the moderate UV break found in the more common
class\,A quasars studied by B05. The outer dust screen is
characterized by a column of \NN=4.3. Hence, the {\it total}
absorption column is the same as in the dust absorbed model\,\II,
but the far-UV flux rise is now better reproduced
(Fig.\,\ref{fig:ton}).

\subsubsection{Limits on atomic gas absorption}\label{sec:atom}

It may be that the roll-over in quasar takes place at higher
energies, as suggested by H07, and that the offset of the flux rise
has a different origin or is the result of the limited S/N of the
data. One possibility to consider is absorption by atomic gas, since
dust and gas must coexist within the screen. We determined the
maximum amount of photoelectric absorption by  \HI\ and \HeI\ that
provided an acceptable fit to the UV trough. This is represented by
the cyan dashed line in Fig.\,\ref{fig:ton}. Note that the input
\sed\,\I\ used in this case does not contain any roll-over function
\Cut. The maximum columns of atomic gas  inferred for this
Model\,\I\ with $\NN=5.3$, which still leads to an acceptable fit of
the UV trough, are $N= 4\times 10^{16}$\,\cms\ and $2\times
10^{17}$\,\cms, for \HI\ and \HeI, respectively. The main conclusion
is that  dust models imply that the associated gas must be highly
ionized. The limits on the neutral fraction of either specie depends
on the dust-to-gas ratio, which can easily be 10 times below that
assumed here. On the other hand, the metallicity inferred from
quasars can be an order of magnitude higher (Hamann \& Ferland 1999;
Dhanda et\,al. 2007) than solar, which would compensate for the
effect of using a lower dust-to-gas ratio. For our assumed
dust-to-gas ratio (see Sect.\,\ref{sec:cur}), we derive an upper
limit for the hydrogen neutral fraction of $\le 8 \times 10^{-5}$.
For helium, assuming a relative abundance of 10\%, the limit on the
neutral fraction is $\le 4 \times 10^{-3}$.

Both values are quite small, especially in the case of \HI\ from
which we infer an ionization parameter $\up \simeq 0.05$ (estimated
using the code \map\ and \sed\,\II).
If we were to consider collisional ionization instead of
photoionization, both neutral fractions are attained when the
temperature is in the range 60000--63000\,K. The high ionization of
the screen probably implies that the dust is not in equilibrium with
the radiation field of \ton. We note that photoelectric absorption
by \HeI\ actually improves the fit to the far-UV rise.
Interestingly, the \sed\ from the  best studied quasar for
absorption line purposes, \HE, reveals a significant dip near the
\HeI\ ionization threshold at 504\,\AA\ (Reimers et\,al. 1998;
Binette et\,al. 2007).

\subsubsection{An outflowing ionized absorber?}\label{sec:out}

\begin{figure}
\resizebox{\hsize}{!}{\includegraphics{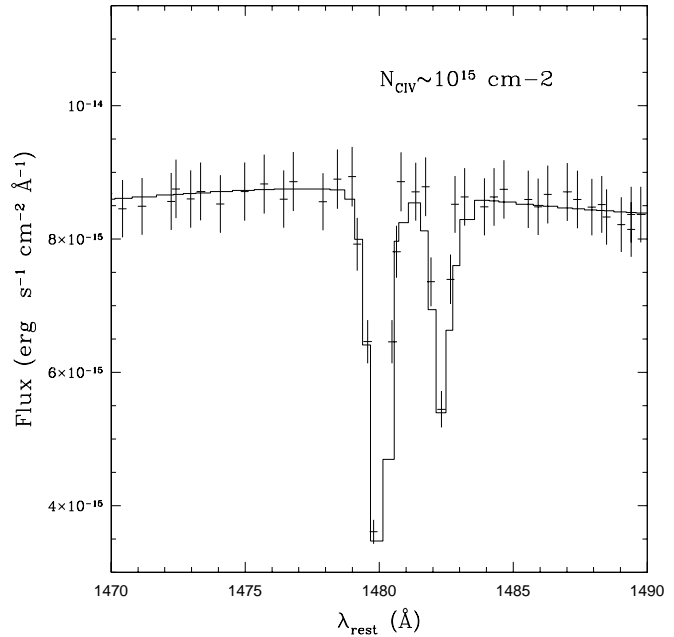}}
\caption{Plot of the \CIV\ absorption system blueshifted by
13200\,\kms. The continuous line is a fit to the \CIV\ doublet
assuming a column of \NCIV=$10^{15}$\,\cms. \label{fig:nciv}}
\end{figure}

The high ionization of the gas required  by the dust model (see
Sect.\,\ref{sec:atom}) leads us to expect the presence of absorption
lines from highly ionized species. In particular, the \CIVw\ line
should be present at some level unless all of C were fully depleted
into dust or overly ionized. We have searched for the presence of
absorption features in the spectra of \ton. We found a significant
\CIV\ absorption system near 1480\,\AA\ (see Fig.\,\ref{fig:nciv}).
Assuming that this absorption system is associated with \ton\ and
contains the crystalline dust responsible for the UV trough, we
derive an outflow velocity of 13200\,\kms, and after fitting the
\CIV\ doublet (solid line in Fig.\,\ref{fig:nciv}), we estimate the
column density to be $\NCIV \sim10^{15}$\,\cms.

\begin{figure}
\resizebox{\hsize}{!}{\includegraphics{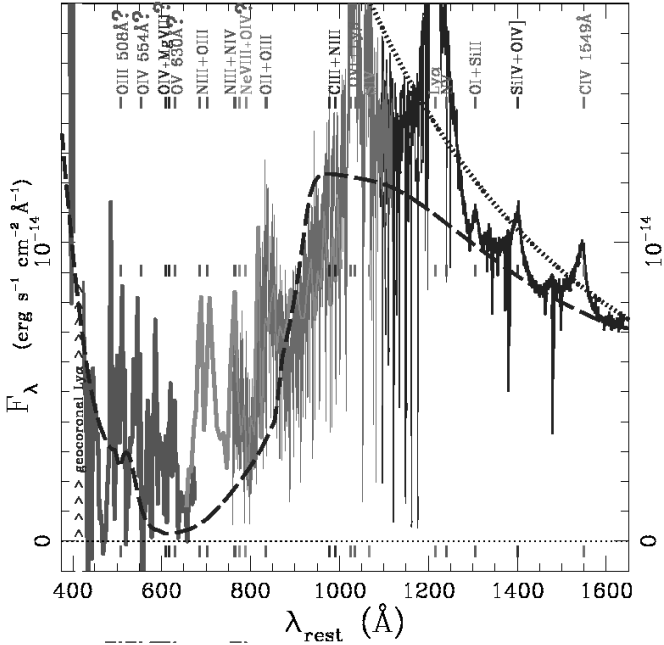}}
\caption{{Same UV energy distribution as in Fig.\,\ref{fig:ton}
(grey scale). The dotted line corresponds to \sed\,\II\ and the
dashed line represents this \sed\ absorbed by nanodiamond dust
(\NN=5.8). The dust is assumed to be present in the \CIV\ absorption
system that we interpret as outflowing  at a  velocity of
13200\,\kms\ (see Sect.\,\ref{sec:out} and Fig.\,\ref{fig:nciv}).
The dust grain size distribution in this model extends from 3 to
100\AA. Absorption by atomic \HI\ is also included, assuming the
same column as derived in Sect.\,\ref{sec:atom} (i.e. $4\times
10^{16}$\,\cms). Fiducial marks indicate the position of observed or
expected (labeled with symbol '?') emission lines. }
\label{fig:out}}
\end{figure}

In Sect.\,\ref{sec:pow} we assumed that all of C was in crystalline
form for the sole purpose of procuring a convenient normalization.
This resulted in an H dust screen column $\NN=5.3$.  The total
column of C that need to be ''depleted onto dust'' in the dust
Models \I\--\III\ presented in Fig.\,\ref{fig:ton} is equivalent to
$\NCdust\sim1.9\times10^{17}$\,\cms. We now explore different
scenarios concerning metallicity and depletion that can provide the
required dust column.

If we adopt the same photoionization model quoted in
Sect.\,\ref{sec:atom}, which is consistent with our estimated limit
on \HI, we infer\footnote{In this same model, the relative fractions
of species \CV/C, \NV/N and \OVI/O are 72\%, 24\%  and 14\%
respectively.} that 13\% of gaseous C is in the form of \CIV\ and,
therefore, the total column of {\it gaseous} C present is
$\NCgas\sim7.7\times10^{15}$\,\cms, which is very small by
comparison with the amount needed in crystalline form. The {\it
total} carbon column (gas $+$ dust) becomes
$\NC\sim2.0\times10^{17}$\,\cms, which translates into a metallicity
only 3\% higher than solar. The depletion of C onto dust in this
case is $\dc=0.96$. A smaller depletion fraction or a higher
ionization parameter would both require larger metallicities. For
instance, a much smaller depletion fraction of 0.1 and a \CIV/C
ratio of 0.13 or 0.013 would imply metallicities of 4.6 and 9 times
solar, respectively\footnote{For a fixed \NCdust, the total column
\NN\ implied scales inversely with metallicity and with \dc.}. We
note that the environment of quasars is often characterized by much
larger than solar metallicities (e.g. Dhanda et\,al. 2007; Hamann \&
Ferland 1999).

Absorption lines of \NVw\ and \OVIw\ are also observed at the same
velocity shift. In the case of \OVI, the doublet is severely blended
with  narrow \Lya\ lines, which prevents us from estimating its
column. As for \NV,  the column is estimated at $7 \times
10^{14}$\,\cms, which is much smaller than that expected from the
$U=0.05$ model, assuming solar metallicity. A much higher ionization
parameter is thus favored by \NV. Finally, a strong \Lya\ absorption
line is also found at the outflow velocity of 13200\,\kms. The line
is slightly saturated, allowing us only to derive  a lower limit to
the \HI\ column of $\simeq 2 \times 10^{15}$\cms, consistent with
the upper limit on allowed \HI\ in Sect.\,\ref{sec:atom} of
$\NHI=4\times10^{16}$. Increasing (decreasing) the ionization
parameter would decrease (increase) this upper limit.

When the dust model is shifted in velocity,  the fit to the far-UV
break worsen significantly. It results into an unacceptable flux
excess at the onset of the UV trough near 950\,\AA. The dust model
can be improved and made to fit both extremities by broadening the
grain size distribution, although by half as much as proposed by
H07. We show in Fig.\,\ref{fig:out} our best fit assuming \sed\,\II\
and based on a grain size distribution that extends from 3 to
100\,\AA. The dust screen contains 10\% more dust than previous
models (i.e. $\NCdust\sim2.1\times10^{17}$\,\cms) and the above
metallicity estimates must be scaled accordingly. Overall, the fit
is quite acceptable.

We note that the outflow velocity of the absorbing material is large
compared to that observed in Narrow Absorption Line Systems found in
Seyfert galaxies (Crenshaw et\,al. 2003). However, \ton\ is much
brighter than those objects. In addition, this system seems somewhat
strong to be part of the IGM, as the metallicity of the \Lya\ forest
at redshifts similar to that of \ton\ is at least 2 order of
magnitude smaller than solar (e.g. Pettini 2006). Thus, it is
certainly possible that both the ionized absorption lines and the
spectral shape in the spectra of \ton\ are produced by the same
screen of material. This would require that the dust is part of a
large scale outflow, as suggested in Sect.\,\ref{sec:disc}.

\section{Discussion on the crystallite dust model}\label{sec:disc}

The \sed\ observed in  \ton\ is rather unique, due to a striking
lack of soft ionizing photons. The far-UV break is definitely more
extreme than in the other quasars of the TZ02 sample (B05). There is
no consensus on the physical origin of the quasar far-UV break.
Binette et\,al. (2007) have recently presented a review of
alternative interpretations, some of which, however, lack detailed
calculations that prevent making a direct comparison with
observations. Among the promising interpretations lies the
possibility of continuum reprocessing by a  wind arising from the
accretion disk or absorption from gas progressively accelerated up
to quasi relativistic velocities (Eastman, MacAlpine \& Richstone
1983). In this paper, we explore the dust absorption interpretation,
for which we dispose of a detailed model, but it is certainly
premature to exclude  other interpretations at this stage.

The nanodiamond dust absorption hypothesis is consistent with the
\ton\ \sed, since it produces a broad absorption feature that is
compatible with the position, depth and shape of the observed
trough. However, the far UV-rise takes place at slightly shorter
wavelengths than predicted by the simplest model (Model\,\I). If not
due to limitations in the S/N or systematics in the instrumental
system of IUE, this can be accounted for by assuming a roll-over
near 670\,\AA\ or photoelectric absorption by \HeI. Alternatively,
the shortcoming might result from the limitations of the dust grain
model. Improved quality and higher resolution spectra will be
essential to determine the true continuum level at the bottom of the
trough and to confirm whether  a very steep rise occurs in the
extreme UV.

Reproducing the observed  continuum trough, in any case, is not
sufficient to vindicate the dust model. For instance, we need to
find evidence in the mid-IR of re-radiation  of the energy absorbed
by the dust. Using Spitzer data on 3C298, de\,Diego et\,al. (2007)
ruled out the presence of emission bands at 3.43 and 3.53\mic\ as is
expected in the case of meteoritic nanodiamonds [i.e. nanodiamonds
with surface adsorbates that produce C$-$H stretch emission
(Andersen et\,al. 1998; Jones \& d'Hendecourt 2000)]. The model used
in this paper is not affected, since it is based on cubic diamonds,
which may emit either via the one-phonon mode due to internal bulk
impurities within the crystals (Andersen 1999)  or, if made of pure
crystals, via multi-phonon modes, which are unfortunately quite
inefficient (Braatz et\,al. 2000; Edwards 1985; Jones \&
d'Hendecourt 2000). However, testing these two remaining
possibilities still remains a task to be done.

The absence of strong bound-free absorption of \HI\ shortward of the
Lyman limit implies a high ionization parameter ($\up \ga 0.05$) for
the dust screen, if we assume photoionization  (or a temperature
above 63\,000\,K for the thermally ionized case). In such an
environment, crystallite dust would not lie in thermal equilibrium
with the ambient gas and therefore must continually be replenished
at a rate near that at which it is being destroyed. A dust screen
within a thin funnel-shaped wind structure as contemplated for the
ionized gas by Elvis (2000) and as inferred from the warm absorbers
in NGC\,4051 (Krongold et\,al. 2007) might resolve this problem. Two
conditions would have to be met, however: the wind must be launched
from a cool region of the accretion disk where the dust can be
formed (see Konigl \& Kartje 1994; Everett 2005) and the wind must
cross our line-of-sight to the UV emission region of the disk.

The possibility of such a wind is particularly relevant to the
problem of detecting the expected absorption lines (such as \CIV\ or
\OVI)  associated with the dust. We do find such lines in an
absorption system that is blueshifted by 13200\,\kms. To validate
this possibility, it is necessary to confirm the presence of similar
absorption systems in other quasars that possess a prominent far-UV
break.

To conclude, despite the success of the  crystalline dust model in
reproducing reasonably well the  far-UV trough observed in \ton\ and
in other quasars (BI05), there arise various objections to such
models that we have partially addressed in this paper. Further
studies are needed to reach firm conclusions in favor of or against
the possibility of crystalline dust as an explanation of the far-UV
break in quasars.


\begin{acknowledgements}
This work was supported by the CONACyT grants J-50296 and J-49594,
and the UNAM PAPIIT grant IN118905.
Diethild Starkmeth helped us with proofreading.
\end{acknowledgements}


\begin{appendix}
\section{Comparison of different dust models}\label{app:a}


\subsection{How grey is the UV extinction?}

There have been many interesting discussions in the literature about
which extinction curve is the most appropriate for active galactic
nuclei (AGN). Following an analysis of 72 optical spectra (of Baker
\& Hunstead 1995) of FR\,II radio-quasars and broad-line
radio-galaxies, Gaskell et\,al. (2004) derived an extinction curve,
which is essentially flat ('grey') shortward of $3800$\,\AA.
However, Crenshaw et\,al. (2001), Cerny et\,al. (2004), Gaskell \&
Benker (2007), Richards et\,al. (2003) and Willott et\,al. (2005)
found an extinction that is increasing towards the UV, at least down
to 1200\,\AA, and maybe beyond. These studies tend to agree on that
there is little or no evidence of the absorption feature near
2175\,\AA\ (see review by Li 2007), a feature that  is otherwise
striking in the case of Galactic extinction. The presence of a
significant amount of scattered light in the particular case of
radio-galaxies may be related to the greyness of the extinction,
which was inferred by Gaskell et\,al. from the Baker \& Hunstead
sample. In their spectro-polarimetric study, Vernet et\,al. (2001)
concluded that the scattering efficiency of quasar light within
their sample of high-redshift radio-galaxies is essentially grey.
The explanation given is that the condensations that cause the
scattering are opaque throughout the UV domaine. In this case, the
scattering efficiency depends only on the albedo, not on selective
extinction. As it turns out, the albedo of Galactic dust is
approximately grey except near the 2175\,\AA\ feature.

\subsection{Multi-component dust models and accretion disks}

Most studies find evidence of an extinction that is rising towards
the far-UV in type\,I AGN (i.e., excluding radio-galaxies). There
are no firm conclusions yet about where in the UV the extinction
becomes grey and starts to decline. It is likely that the dust
grains vary in composition and optical properties according to the
particular line-of-sight in which the observer happens to lie.
Furthermore, there can be more than one dust component present in
any given line of sight. For instance, Gaskell \& Benker (2007)
suggest an extinction similar to the Galactic (although without the
usual 2175\,\AA\ absorption feature), but accompanied, at least in
some of their AGN, by additional extinction due to a SMC-like dust
component. The possible presence of more than one dust component is
the starting point of our attempt to account for the far-UV break of
quasars, for which we will assume it is due to an additional dust
component. We used an inverse technique, however, by exploring
different dust compositions and size distributions, until the
extinction curve we calculated could account for the very sharp
break observed near 1100\,\AA.

We emphasize that we are not questioning the general belief that the
BBB corresponds to emission from an accretion disk around a
supermassive blackhole. A persistent  problem, however, is that the
observed BBB is much too soft to account for the high excitation
emission-lines typical of quasars (KO97; Koratkar \& Blaes 1999).
Our aim has been to explore alternative solutions. Some of these
have been ruled out, while others still lack detailed calculation to
enable proper testing  (see review by Binette et\,al. 2007). In our
opinion, either there exists a secondary ionizing continuum
component in the (unobserved) extreme UV (KO97; Binette, Courvoisier
\& Robinson 1988), or the observed  steepening of the UV continuum
is a relatively narrow feature followed by a continuum recovery at
higher energies (this Paper). Since all dust models reach an
absorption peak somewhere in the near or far-UV, followed by a
decrease in cross-section towards the soft X-rays, the dust
hypothesis always implies a continuum recovery towards the extreme
UV.

\subsection{The far UV-break in \ton\ and various dust models}

From an early exploration of various dust models B05 concluded that
crystalline nanodiamond dust had the required optical properties to
reproduce the far-UV break of quasars. The proposed dust model is
based on a relatively abundant element, namely carbon. We compare
 this with alternative dust models below.

\begin{figure}
\resizebox{\hsize}{!}{\includegraphics{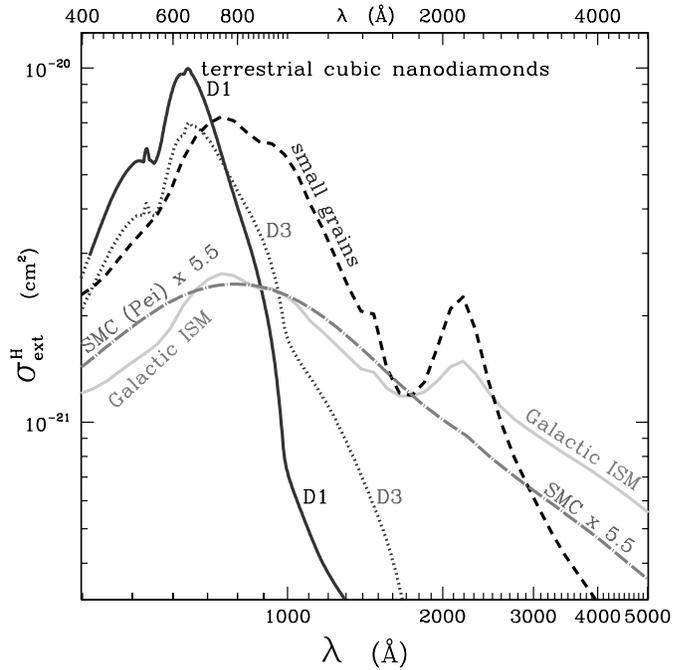}}
\caption{Comparison of the extinction cross sections from different
dust models: black continuous line: (D1) nanodiamonds of sizes $\le
25$\,\AA\ (see Sect.\,\ref{sec:cur}); dotted line: (D3) nanodiamonds
of sizes extending up to 200\,\AA\ (H07); continuous silver line:
model by Martin \& Rouleau (1991) of the Galactic extinction
consisting of graphite and silicates grains of sizes between 2500
and 50\,\AA; dashed line: extinction by graphite and silicate grains
of small sizes $\le 300$\,\AA; long-dash-dotted line: the SMC dust
extinction model of Pei (1992) multiplied by a factor 5.5 to
facilitate comparison. \label{fig:mod}}
\end{figure}

\begin{figure}
\resizebox{\hsize}{!}{\includegraphics{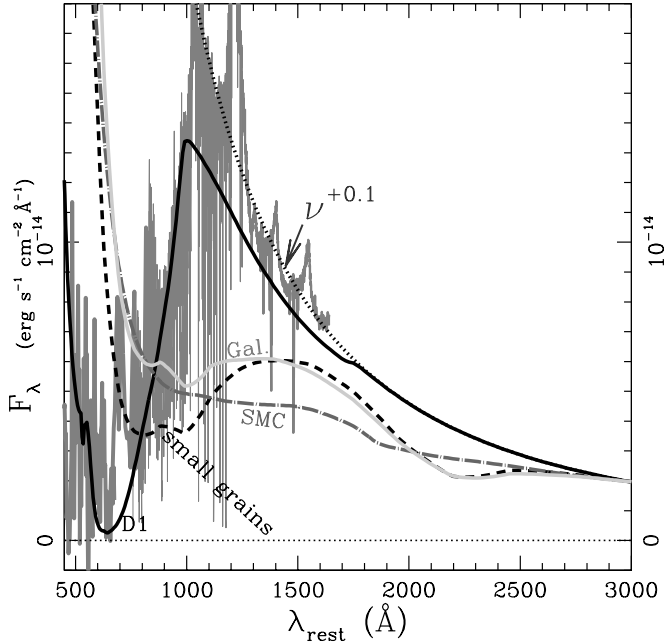}}
\caption{Comparison of the extinction that results from  various
dust models (c.f. Fig.\,\ref{fig:mod}).  The assumed \sed\ is a
powerlaw of index +0.1 (dotted line). The \ton\ spectrum of (see
Fig.\,\ref{fig:ton}) is overlayed (thin grey line). To facilitate
comparison, {\it all} curves shown were normalized to the same flux
at 3000\,\AA. The curves are labeled as follows: black continuous
line:  nanodiamonds in small size regime (D1); continuous silver
line: model by Martin \& Rouleau (1991) of the Galactic ISM
extinction; dashed line: small graphite and silicate grains;
long-dash-dotted line: SMC dust extinction (SMC dust model from
H07). \label{fig:comp}}
\end{figure}

In Fig.\,\ref{fig:mod} we show the extinction curve D1 used in the
current paper (black continuous line). For all models shown below,
we assume a powerlaw size distribution ($\propto a^{-3.5}$, see
Mathis et\,al. 1977; Draine \& Lee 1984). The curve D1 illustrates
the small size regime ($a \le 25$\,\AA) for nanodiamonds, while
curve D3 is characterized by grain sizes spanning a wider range of
$3 \le a \le 200$\,\AA. The \sed\ steepening in Class\,B spectra
(defined in B05) such as in \ton\ is well fitted using  dust model
D1 while the more numerous Class\,A spectra appear to favor
extinction model D3 (see H07). In Fig.\,\ref{fig:comp}), we show a
simple powerlaw absorbed by a dust screen of column \NN =5.3
characterized by an extinction given by model D1.

We now compare the extinction resulting from other dust models shown
in Fig.\,\ref{fig:mod}. The silver line corresponds to a model of
the Galactic extinction as calculated by Martin \& Rouleau (1991).
It contains an equal number of graphite and silicate grains with
grain sizes encompassing the range $50 \le a \le 2500$\,\AA. A
comparison with the \ton\ is presented in Fig.\,\ref{fig:comp}
assuming a column \NN=10. Another dust model is represented by the
black dashed line, which is similar to the previous, but with a size
range confined to small grains in the range $50 \le a \le 300$\,\AA.
The absorbed \sed\ in Fig.\,\ref{fig:comp} corresponds to \NN=3.4.
The last extinction  considered is a model of the SMC extinction
consisting of  amorphous carbon grains with  grain sizes comprised
within the limits $50 \le a \le 1400$, as calculated by H07. The
dust column assumed in Fig.\,\ref{fig:comp} is \NN=10.

The dust columns that we have assumed are arbitrary, but suffice to
illustrate the contrasting behavior of the absorption in the far-UV
that the various dust models produce. An inspection of
Fig.\,\ref{fig:comp} shows that only nanodiamonds have the potential
to reproduce the far-UV break observed in \ton.

Although dust in crystalline form is not common place in the Galaxy
(Whittet 2002), this may not be the case in AGN. For instance, Roche
et\,al. (2007) report the detection of a spectral structure near
11.2\mic\ in NGC\,3094, indicative of the possible presence of
crystalline silicates.  The presence of an ultraviolet radiation
field might favor the formation of nanodiamonds through one of the
following processes: UV annealing of carbonaceous grains (Nuth \&
Allen 1992), nucleation in organic ice mixtures by UV photolysis
(Kouchi et\,al. 2005) and chemical conversion of PAH clusters to
nanodiamonds  (Duley \& Grishko 2001). The absence of silicates
could be explained by its significantly lower sublimation
temperature. The resilience of small nanodiamonds led Rouan et\,al.
(2004) to favor these as  candidates to explain the IR emission from
the 4 elongated nodules that they spatially resolve (K, L and M
bands) in the nucleus of NGC\,1068.

\end{appendix}

\end{document}